\RequirePackage{amsmath}

\documentclass[a4paper]{article}       % onecolumn (second format)
\usepackage[utf8]{inputenc}
\usepackage[T1]{fontenc}
\usepackage{graphicx}
\usepackage{colortbl}
\usepackage[vlined,algoruled]{algorithm2e}
\usepackage{algorithmic}
\usepackage[english]{babel}
\usepackage{amsmath, amsthm}
\usepackage{thm-restate}
\usepackage{tikz}
\usepackage{enumerate}
\usepackage{wrapfig}
\usetikzlibrary{decorations.pathreplacing,calc,positioning,decorations.markings}
\usepackage{caption, subcaption}
\usepackage {nicefrac}
\theoremstyle{definition}
\newtheorem{definition}{Definition}[section]
\theoremstyle{plain}
\newtheorem{theorem}[definition]{Theorem}

\theoremstyle{remark}

\usepackage{url}
\usepackage{xcolor}
\usepackage{txfonts}
\usepackage{todonotes}
\usepackage{float}

\usepackage{tkz-graph}
\usepackage{widetable}
\usepackage{booktabs}
\usepackage{natbib}

\newcommand{\sectionheadline}[1]{\paragraph{\tb{#1.}}}
\newcommand{\tb}[1]{\textbf{#1}}

\newcommand{\N}{\mathbb{N}}

\newcommand{\R}{\mathbb{R}}

\newcommand{\val}{\text{Val}}

\newcommand{\stackcell}[2]{\begin{tabular}{@{}c@{}}#1\\#2\end{tabular}}

\newcommand{\Halmos}{$\square$}

\title{Computational Methods for Path-based Robust Flows}
\date{\today}
\author{Fabian Mies\thanks{Institute of Statistics, RWTH Aachen University, {mies@stochastik.rwth-aachen.de}} \and Britta Peis\thanks{School of Business and Economics, RWTH Aachen University, {peis@oms.rwth-aachen.de}} \and Andreas Wierz\thanks{School of Business and Economics, RWTH Aachen University, {wierz@oms.rwth-aachen.de}}}

\begin{document}

\maketitle

\begin{abstract}
Real world networks are often subject to severe uncertainties which need to be addressed by any reliable prescriptive model.
In the context of the maximum flow problem subject to arc failure, robust models have gained particular attention.
For a path-based model, the resulting optimization problem is assumed to be difficult in the literature, yet the complexity status is widely unknown.
We present a computational approach to solve the robust flow problem to optimality by simultaneous primal and dual separation, the practical efficacy of which is shown by a computational study.

Furthermore, we introduce a novel model of robust flows which provides a compromise between stochastic and robust optimization by assigning probabilities to groups of scenarios.
The new model can be solved by the same computational techniques as the robust model.
A bound on the generalization error is proven for the case that the probabilities are determined empirically.
The suggested model as well as the computational approach extend to linear optimization problems more general than robust flows.

\end{abstract}

\section{Introduction}
Network flow problems are among the most studied topics in combinatorial optimization, computer science and operations research.
In many variants, the complexity status is entirely understood and often very efficient algorithms exist.
In this work, we consider a special network flow problem, namely the maximum flow problem, which seeks to maximize the total throughput in a capacitated network.
Although it has first been studied by Ford and Fulkerson in the 1950s, new, faster algorithms still appear.
The latest contribution of this kind was only four years ago and is due to \cite{orlin2013max}.

The large interest in network flow problems stems from at least two reasons.
Network structures appear in a large range of different problems.
Thus, a better understanding of the polyhedral structure of network flow problems lead to efficient algorithms for adjoint problems in the past.
As an example, the linear relaxation of the precedence constrained knapsack problem can be solved by means of network flow problems very efficiently.
This has a direct application to open pit mining (\cite{moreno2010large}).
But they have broad, direct applications in practice as well, such as production planning, scheduling, and logistics.
Such applications usually are accompanied by data uncertainty.
Shipments of parts in a production environment may be delayed, thus, reducing short-term production capacities.
Means of transport in a logistics network may be delayed or attacked, thus, resulting in a reduction of the total throughput of particular links.
Classical results regarding network flow problems do not take such issues into account but require precise input data.
Moreover, they may compute solutions which are optimal for precise data, but may perform bad or turn infeasible after data slightly changed.
Hence, the urge for robust solutions is constantly rising.

Network flow problems under uncertainty have been studied in a range of different settings.
The structure of the uncertainty set and the demanded type of solution depend on the actual application and influence the problem complexity.
Here, we consider the problem in the robust optimization framework as suggested by \cite{BeSi03}, denoted as the $k$-robust flow problem.
Given a graph $G = (V,E)$ with designated vertices $s$ and $t$, we want to find an $s$-$t$-flow which has maximum value after the removal of at most $k \in \N$ edges.
In this setting, the flow may not adapt to the set of edges being destroyed but has to be fixed beforehand.
Any flow shipped on paths intersecting the set of removed edges is destroyed.
The set of possible scenarios can be described as the subsets of edges of cardinality at most $k$.

Adaptive optimization models for problems under uncertainty have also been studied in the past.
In such models, the solution may adapt to each of the scenarios individually in order to improve the solution quality.
Since the set of scenarios can be of exponential size, the adaptation is usually not provided explicitly.
Instead, the solution is accompanied by a recovery algorithm which computes the adaptation on demand.
Although such models can provide better solution quality, many practical applications do not support such short-term adaptations.
Capacities in production or logistics networks are usually planned and reserved in advance and changes take time in order to be propagated properly.
Hence, we consider the robust, more restrictive, model in most of this work.

In practical applications, the worst-case may only appear with a very small probability.
In such situations, evaluation of solutions with respect to their worst-case performance may be too pessimistic.
Instead, stochastic optimization methods may be applied.
In case of maximum flows, one could maximize the expected flow value given a probability distribution for the joint failure of any set of edges.
Unfortunately, detailed stochastic models are computationally intractable in most cases.
Moreover, specification of the full probability distribution is unreliable since observation of historical events does not necessarily yield good estimates on the probability of events that never appeared before.
In Section \ref{sec:stochastic-model}, we consider a model which remedies both aspects by aggregating the occurrence of single events into larger classes.

\sectionheadline{Related work}

\cite{wood1993deterministic} showed that the network interdiction problem is NP-hard.
The network interdiction problem is related to the robust flow problem with the order of optimization changed.
Network interdiction minimizes the maximum flow value of a network, instead of maximizing the minimum flow value in the worst-case. For this problem, \cite{altner2010maximum} provide valid inequalities and prove integrality gaps for natural LP formulations. Approximability of the network interdiction problem is studied, for example, by \cite{chestnut2017hardness}.

\cite{aneja2001maximizing} show how the $k$-robust flow problem (robust flow problem in the following) can be solved for $k = 1$.
For $k = 2$, \cite{du2007maximum} showed that the dual separation problem to the $k$-robust flow model we introduce in Section \ref{sec:robust-flow-model} is NP-hard.
They reason that - due to the equivalence of optimization and separation - this implies that the optimization problem is also NP-hard.
\cite{disser2017complexity} pointed out that this reasoning is flawed.
Du and Chandrasekaran proved hardness for arbitrary objective functions, whereas the objective function stemming from the robust flow LP is very specific.
Hence, the equivalence does not apply and the complexity status remains open, if $k$ is bounded by a constant greater than one.
\cite{disser2017complexity} showed that the problem is NP-hard, if $k$ is not bounded.

A number of approximation results were shown for robust flow problems. \cite{bertsimas2013robust} present an approximation algorithm for $k$-robust flows whose performance may be bounded in terms of the optimal solution. The same algorithm can be shown to achieve the approximation bound $(k+1)/(k^2/4+k+1)$.
As an alternative approach, \cite{baffier2014k+} study $k$-route flows (see also \cite{aggarwal2002multiroute,kishimoto1992m,kishimoto1996method}). A $k$-route flow is a conic combination of elementary flows which in turn consist of $k$ disjoint paths. \cite{baffier2014k+} showed that a $k$-route flow can be used to obtain a $(k+1)$-approximation for $k$-robust flow and evaluate the practical applicability of their theoretical results (\cite{baffier2015algorithms}).
\cite{baffier2014parametric} also showed that the algorithm sometimes determines an optimum solution to the problem and provide conditions under which this can be checked a posteriori. We note that the approximation bound derived by \cite{bertsimas2016power} is tighter than the latter.

Various other approaches to study network flows subject to arc failures have been suggested in the literature. \cite{matuschke2015protection} showed that the robust flow problem becomes tractable if the model is altered such that, instead of edges, specific paths are attacked.
In this case, they show that the problem can be solved in polynomial time. They also consider an extension of the model where each arc may be fortified against attacks by paying an arc-specific cost.
Aneja et al.\ discuss so-called $\delta$-reliable flows.
A flow is said to be $\delta$-reliable, if no edge carries more than a $\delta$ fraction of the total flow value (\cite{aneja2007flows,baffier2014parametric}).
Such solutions can be computed in polynomial time. \cite{bertsimas2013robust} suggest an arc-based flow model which allows for an adaptation once an interdiction has occurred and is closely related to the network interdiction problem. A different adaptation scheme is studied by \cite{matuschke2017rerouting}, who introduce a path-based model which only allows for local adaptations of the scheduled flow.  

\cite{gottschalk2016robust} discuss the $k$-robust flow problem with an additional temporal component.
They provide insight regarding the power of temporally repeated solutions.

\sectionheadline{Our contribution}

In this paper, we perform a computational study which indicates that, in practice, robust flow problems are tractable in many situations.
We show that a natural LP formulation with exponentially many constraints and variables can be solved in reasonable time using simultaneous primal and dual separation techniques.
This is partially due to the insight that the robust flow problem admits a sparse solution for various types of networks.

On the theoretical side, we introduce a variant of $k$-robust flow weighting different values of $k$.
The new model may be treated by the presented computational approach as well.
Furthermore, we state the precise relation of the weighted model to stochastic and robust optimization, generalizing a result of \cite{bertsimas2016power}.
Our result applies to more general optimization problems under uncertainty than robust flows and presents a compromise between detail and robustness of the description of an uncertain setting.
By employing a Rademacher complexity bound, we show that assessing the performance of a flow solution based on historical failure scenarios is more reliable for robust flows than for the solution of a fully stochastic model.

\sectionheadline{Outline}

The rest of this paper is organized as follows.
In Section \ref{sec:robust-flow-model}, we introduce our model for robust flows. In Section \ref{sec:solution-techniques}, a separation procedure is described to obtain primal and dual bounds on the robust flow value.
Computational results are presented in Section \ref{sec:computational-results}.
In Section \ref{sec:stochastic-model}, we discuss a connection between a stochastic optimization model for flows under uncertainty and the robust model from Section \ref{sec:robust-flow-model}.
We introduce a hybrid model which subsumes both problems at its extremes.
The paper is concluded with a summary of the results.

\section{Robust flow model}
\label{sec:robust-flow-model}

\sectionheadline{Nominal maximum flow problem}

\cite{ford1956maximal} were the first to study the maximum flow problem.
They provided the famous and influential max flow - min cut theorem 60 years ago.
Although the problem is usually known in its edge based formulation, the first appearance was stated in terms of a path formulation.
Given a graph $G = (V,E)$ with edge capacities and two designated vertices $s,t \in V$.
Find a set of paths with flow rates such that no edge capacity is exceeded and the total throughput is maximized.
Formally, the set of feasible flows can be written as
\begin{align*}
\mathcal{F} = \left\{ \left(x_P\right)_{P\in\mathcal{P}} \; \Big| \; \sum_{\substack{P \in \mathcal{P} :\\ e \in P}} x_P \leq u_e \forall e \in E, \; x \geq 0  \right\},
\end{align*}
where $\mathcal{P}$ denotes the set of all s-t-paths in $G$.
By $x_P$ we denote the flow rate on path $P \in \mathcal{P}$.
The optimization problem is given by
\begin{align*}
\max \left\{ \sum_{P \in \mathcal{P}} x_P \; \big| \; x \in \mathcal{F} \right\}.
\end{align*}

\begin{figure}[tb]
	\centering
	\includegraphics[page=1]{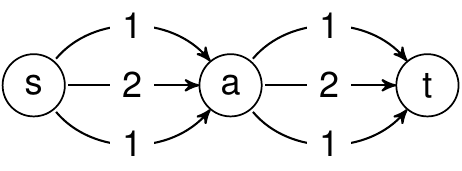} \\
	\includegraphics[page=2]{robust-flow-model-decomposition} \qquad
	\includegraphics[page=3]{robust-flow-model-decomposition}
	\caption{Example graph showing that that the robust flow value depends on the specific path decomposition. Top: arc flow. Bottom left: path decomposition with robust value $1$. Bottom right: path decomposition with robust value $0$.}
	\label{fig:robust-flow-model-decomposition}
\end{figure}

Nowadays, we know a large variety of algorithms which usually solve the problem in terms of the equivalent edge formulation.
While both formulations are equivalent in the nominal case, that is, without any kind of robustness,
the actual path decomposition matters when we consider the robust counterpart.
See Figure \ref{fig:robust-flow-model-decomposition} for an example.

\sectionheadline{Robust maximum flow problem}

In the remainder of this work, we elaborate and evaluate computational techniques which help solving the following robust counterpart of the maximum flow problem.
For any feasible flow vector $x \in \mathcal{F}$ we define the flow value with respect to scenario $\eta \in \binom{E}{k} = \{ C \subseteq E : |C| = k \}$ as the total amount of flow using paths which avoid edges in $\eta$.
That is, $\val_\eta(x) = \sum_{P : P \cap \eta = \emptyset} x_P$.
Moreover, we denote the robust flow value for $x \in \mathcal{F}$ as the flow value with respect to the worst case scenario, that is, $RVal(x) = \min_{\eta \in \binom{E}{k}} \val_{\eta}(x)$.
The robust maximum flow problem seeks to maximize this value.
One way of formulating this problem as a linear program is as follows
\begin{align}
\max \left\{ \sum_{P \in \mathcal{P}} x_P - \lambda \; \big| \; x \in \mathcal{F}, \; \lambda \geq \eta(x) \; \forall \eta \in \binom{E}{k} \right\}, \tag{P} \label{lp:robustfull}
\end{align}
where $\eta(x) = \sum_{P : P \cap \eta \neq \emptyset} x_P$ denotes the amount of flow intersecting edges in $\eta$.
The formulation ensures that $x$ is a feasible flow.
Moreover, the value of $\lambda$ is lower bounded by the amount of flow destroyed by each scenario.
Hence, for a feasible solution, $\lambda$ is always at least as large as the amount of flow destroyed in the worst case.
Since $\lambda$ has a negative objective function coefficient in a maximization problem, it will also attain this value.
Thus, the objective function maximizes the robust flow value.

\cite{aneja2001maximizing} showed how (\ref{lp:robustfull}) can be solved for $k = 1$.
\cite{du2007maximum} claimed that the problem for every $k \geq 2$ is NP-hard.
However, \cite{disser2017complexity} recently showed that the proof is incorrect and that the problem is NP-hard if $k$ is not bounded by any constant.
The result is not yet published and the problem remains open if $k$ is bounded by a constant.
The dual separation problem of (\ref{lp:robustfull}) remains NP-hard for $k \geq 2$ as shown by \cite{du2007maximum}.
\cite{bertsimas2016power} provide a relaxation of (\ref{lp:robustfull}) with an approximation ratio of $(k + 1) / (k^2/4 + k + 1)$.

\section{Solution techniques}
\label{sec:solution-techniques}

Due to its size, model (\ref{lp:robustfull}) can neither be stored in memory nor solved efficiently.
Large scale linear programs with an exponential number of constraints are typically solved by exploiting the specific model structure in order to generate constraints which should be part of the basis.
On the other hand, if the number of variables is large, one may generate constraints for the dual LP instead.
The latter procedure is known as column generation, as new variables are added to the primal model (\cite{desrosiers2005primer}).

The robust flow model (\ref{lp:robustfull}) has both, a large number of variables and constraints.
We suggest to solve this issue by generating the set of variables and constraints simultaneously.
To this end, we first investigate the primal and dual separation problem individually.
Subsequently, we show how both procedures benefit from applying them simultaneously.

\sectionheadline{Primal separation}
The huge number of constraints of (\ref{lp:robustfull}) can be tackled by separating them as required.
That is, we solve the model for a subset $\mathcal{N}'\subseteq \binom{E}{k}$ of constraints and only add scenarios for which the corresponding linear inequality is violated.
The separation problem is equivalent to the network interdiction problem, which may be described by the integer program (cf.\ \cite{bertsimas2013robust})
\begin{align}
\label{lpseparation}\min \quad & \sum_{P\in\mathcal{P}} x_P z_P \\
\nonumber \text{s.t.}\quad & \sum_{e\in P} z_e + z_P \geq 1 \quad && \forall P \in \mathcal{P}\\
\nonumber& \sum_{e\in E} z_e = k \quad && \forall e\in E \\
\nonumber& z_e  \in \{0,1\}, \quad z_P\in [0,1]\quad && \forall e\in E, P\in\mathcal{P}.
\end{align}
Problem (\ref{lpseparation}) can be interpreted as an instance of the weighted Maximum Coverage Problem, which is NP-hard in general (see e.g.\ \cite{ageev1999approximation}).
However, the corresponding linear relaxation admits a relative integrality gap which is bounded by $1-e^{-1}$ (\cite{ageev1999approximation}).
Hence, one might expect a standard Branch-and-Bound solver to perform reasonably well on this problem, especially if the set $\mathcal{P}$ of paths is of low cardinality.
Note that solving (\ref{lp:robustfull}) for the linearly relaxed separation problem (\ref{lpseparation}) yields the heuristic of \cite{bertsimas2013robust}.

\sectionheadline{Dual separation}
In a similar fashion, the large set of variables of (\ref{lp:robustfull}) can be handled by means of column generation.
In each iteration, the model is solved with a subset $\mathcal{P}'\subseteq \mathcal{P}$ of path variables to obtain a feasible dual solution $y=(y_e, y_\eta)$.
The pricing problem, which is the separation problem of the dual LP, is to find a path $P$ which maximizes $\sum_{\eta\cap P \neq \emptyset}y_\eta - \sum_{e\in P} y_e$.
As shown by \cite{du2007maximum}, this problem is NP-hard in general.
Yet, if the vector $(y_\eta)_{\eta}$ is sufficiently sparse, a solution may be obtained by applying a Branch-and-Bound procedure to the following integer program:
\begin{align}
\label{lppricing}\max \quad & 1-\sum_{\eta} z_\eta y_\eta -\sum_e z_e y_e\\
\nonumber \text{s.t.}\quad & \sum_{e \in \delta^-(v)} z_e - \sum_{e \in \delta^+(v)} z_e = 0 \quad && \forall v\in V\setminus\{s,t\}\\
\nonumber& \sum_{e \in \delta^-(t)} z_e - \sum_{e \in \delta^+(t)} z_e = 1 \quad \\
\nonumber&  z_e \leq z_\eta \quad && \forall \eta\in\binom{E}{k}, e\in \eta \\
\nonumber& z_e,z_\eta \in \{0,1\}.
\end{align}
If the value of (\ref{lppricing}) is positive, the path encoded by the variables $z_e$ is added to the set $\mathcal{P}'$ and model (\ref{lp:robustfull}) is solved again.
When solving (\ref{lppricing}), only those variables $z_\eta$ with $y_\eta\neq 0$ need to be considered.
Thus, sparsity of $y$ simplifies the pricing problem.

\sectionheadline{Simultaneous separation}
\label{subsec:mainloop}
While the primal separation (\ref{lpseparation}) and dual separation (\ref{lppricing}) problems are both NP-hard, they benefit from a small set of paths and scenarios, respectively.
When generating variables and constraints simultaneously, we maintain a set $\mathcal{P}'\subseteq \mathcal{P}$ of paths and a set $\mathcal{N}'\subseteq \binom{E}{k}$ of scenarios and solve the restricted linear program
\begin{align}
\tag{P'}\label{lprobustsmall}\max \quad & \sum_{P\in\mathcal{P}'} x_P - \lambda \\
\nonumber \text{s.t.}\quad & \sum_{P: e\in P} x_P \leq u_e \quad && \forall e\in E\\
\nonumber& \sum_{P:P\cap \eta\neq\emptyset} x_P \leq \lambda \quad && \forall \eta\in \mathcal{N}'\\
\nonumber& x_P \geq 0 \quad && P\in \mathcal{P}'.
\end{align}
If $\mathcal{P}'$ and $\mathcal{N}'$ are small, (\ref{lprobustsmall}) can be solved fast by a standard LP solver.
The dual solution $y$ may be interpreted as setting $y_\eta=0$ for all $\eta\notin\mathcal{N}'$.
Due to this sparsity, the pricing problem (\ref{lppricing}) may be solved efficiently for the restricted set of scenarios $\mathcal{N}'$, generating a new path which is added to the set $\mathcal{P}'$ of candidate paths.
We keep adding paths until optimality is achieved.
At this point, model (\ref{lprobustsmall}) with optimal value $f^d$ is a relaxation of the full robust flow model (\ref{lp:robustfull}), and $f^d$ is a dual bound on the optimal value $f^*$ of the full model.

For this set $\mathcal{P}'$ of paths, keep adding scenarios by means of the (restricted) separation problem (\ref{lpseparation}) until it no longer affects the solution $(x, \lambda)$ of model (\ref{lprobustsmall}).
Then $(x,\lambda)$ is feasible for the full model.
Thus, the corresponding value $f^p=(\sum_P x_P - \lambda)$ is a primal bound on $f^*$.

Iterating these two steps, the model yields tighter bounds as additional variables and constraints are considered.
Once the primal and dual bounds coincide, that is, $f^p=f^d$, the corresponding flow $f$ is known to be optimal for (\ref{lp:robustfull}).
Since the number of potential variables and constraints is finite, the algorithm eventually terminates.
Empirical results indicate that it terminates after a small number of iterations in practice.

\sectionheadline{Practical aspects}

Once the variables $z_\eta$ are fixed, the dual separation problem (\ref{lppricing}) reduces to a shortest path problem.
Thus, if the corresponding price vector $(y_\eta)_\eta$ is very sparse, an optimal path could be determined by enumerating all configurations of $z_\eta$ and computing the corresponding shortest path in a reduced network.
A similar strategy is realized by only branching on the binary variables $z_\eta$, since (\ref{lppricing}) is totally unimodular for fixed $z_\eta$.

When performing the primal and dual separations, it is not necessary to find an optimal solution to the separation problem at each step as long as it yields a violated constraint.
Therefore, one could try to tweak the procedure by slightly modifying the separation procedures.
In our computational experiments, for the generation of new path variables, we add a small penalty term $\epsilon=10^{-4}$ for each edge which has already been considered by a generated path.
For the interdiction problem, we add a penalty of $1/k$ for each arc which is interdicted by a scenario generated previously.
If the perturbed problem does not yield a violated constraint, we evoke the unperturbed model.
The intuition motivating these modifications is to reduce correlation of the generated constraints.

For very large networks, the required number of primal constraints to solve the primal separation problem could be very large.
Numerical experiments reveal that this issue arises in particular for large values of $k$ (see Section \ref{sec:computational-results}).
As a consequence, each iteration of the simultaneous separation procedure is slowed down, leading to stagnant updates of the primal and dual bounds $\lambda^p, \lambda^d$.
As a remedy, we suggest to generate at most a fixed number $c$ of primal constraints in each iteration.
Each single call of the interdiction LP (\ref{lpseparation}) yields the robust value of a feasible solution, that is, a primal bound which might improve the current bound.
Subsequently, generating paths until optimality for the reduced set of scenarios is achieved restores a dual bound, as described previously.
This iteration scheme leads to primal and dual bounds which narrow the optimality gap gradually.

A suitable initialization of the sets $\mathcal{P}'$ and $\mathcal{N}'$ may help in solving problem (\ref{lprobustsmall}).
The initial set of paths may be generated by means of the heuristic of \cite{bertsimas2013robust}.
They suggest to find a feasible flow $x\in \mathcal{F}$ which maximizes
\begin{align}
\tag{H}\label{lp:heuristic}\sum_{P\in\mathcal{P}} x_P - k\max_e \sum_{\substack{P \in \mathcal{P}:\\ e\in P}} x_P = |x| - k \max_e x_e
\end{align}
where $x_e=\sum_{P\in\mathcal{P}: e\in P} x_P$ is the arc flow corresponding to $x$, and $|x|=\sum_{P} x_P$.
Problem (\ref{lp:heuristic}) can be solved efficiently using standard maximum flow techniques and binary search.
It is shown by \cite{bertsimas2013robust} that any path decomposition corresponding to the optimal solution $x^*$ of (\ref{lp:heuristic}) attains the robust value $RVal(x^*)=|x^*|-k\max_e x_e$.
Hence, any such path decomposition is a reasonable initialization for the set $\mathcal{P}'$.

\section{Computational results}
\label{sec:computational-results}

\sectionheadline{Instances}

\begin{table}[bt]
	\centering
	\small
	\begin{tabular}{lccc}
		\toprule
		Network & Parameters & Nodes & Arcs \\ \midrule
		P1 & $M=50, n=20$ & 3 & 1920 \\\midrule[0pt]
		P2 & $n=5, m_0=20$& 5 & 160 (in mean) \\ \midrule[0pt]
		NETGEN-a &  & 256 & 2048  \\ \midrule[0pt]
		NETGEN-b &  & 512 & 4096 \\ \midrule[0pt]
		R-MAT-a & $(a,b,c,d)=(0.5,0.2,0.2,0.1)$ & $1000$ & $2000$ \\ \midrule[0pt]
		R-MAT-b & $(a,b,c,d)=(0.5,0.2,0.2,0.1)$& $1000$ & $5000$ \\ \midrule[0pt]
		R-MAT-c & $(a,b,c,d)=(0.2,0.2,0.4,0.2)$ & $1000$ & $5000$ \\ \bottomrule
	\end{tabular}
	\caption{Summary of test-cases and the corresponding parameterization.}
	\label{tab:instances}
\end{table}

\begin{figure}[tb]
	\centering
	\includegraphics[page=1]{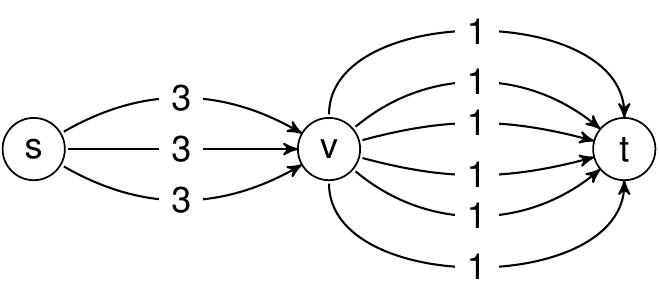}
	\caption{Example of the series graph type P1; $n=3,M=3$.}
	\label{fig:P1}
\end{figure}

The efficacy of the proposed computational decomposition is assessed for several types of networks (see Table \ref{tab:instances}).
Two types of series graphs are considered.
The network P1 consists of three nodes $\{s,v,t\}$ with $n$ parallel arcs $(s,v)$, each with (large) capacity $M\in \mathbb{Z}_+$, and $(n-1)M$ parallel arcs $(v,t)$ with unit capacity (see Figure \ref{fig:P1}).

Furthermore, we generate random series graphs of type P2 with nodes $V=\{v_1,\ldots, v_n\}$, $s=v_1, t=v_n$ and parallel arcs $e^j_1,\ldots, e^j_{m_j}$.
The number $m_j$ of arcs in the $j$-th cut is randomizes, following a shifted Poisson distribution $m_j= m_0 + \text{Poi}(m_0)$.
Capacities are chosen as $u_{e^j_i} = u_{e^j_{i-1}} + (|\epsilon_i^j|+1)^2$ for independent standard normal random number $\epsilon^j_i$.
Thus, the capacities within a single cut are highly correlated, leading to much variation in the capacity of the cuts.

\begin{figure}[tb]
	\centering
	\includegraphics[page=1]{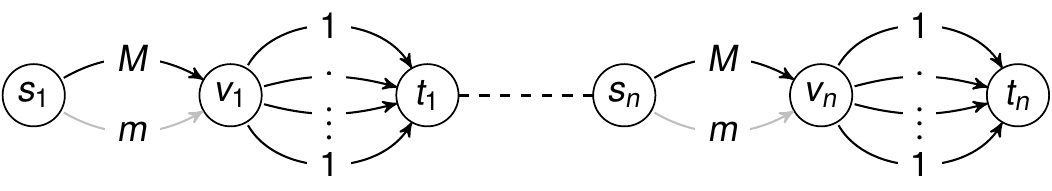}
	\caption{Example of the series graph type P3. The grey edge with capacity $m$ can not be interdicted.}
	\label{fig:P3}
\end{figure}

A third class P3 of series graphs is obtained by connecting a node $s$ to $v$ with a safe arc of capacity $m$ and an interdictable arc of capacity $M$. The node $v$ is connected to $t$ by $m+M$ arcs with capacity $1$. This subgraph is repeated $n$ times to obtain a highly symmetric network of rather small cardinality (see Figure \ref{fig:P3}).

More complex random networks are generated by a model called NETGEN (\cite{klingman1974netgen}).
These instances consist of multiple sources and sinks which may be cast into the framework of s-t-flows by connecting these to a supersource, respectively supersink, by safe arcs.
That is, arcs, which may not be interdicted.
We use the NETGEN-8 instances provided as benchmark by the LEMON graph library project (\cite{dezsHo2011lemon}).

The R-MAT graph generator (\cite{chakrabarti2004r}) is parameterized by four parameters $(a, b, c, d)$ which lead to different types of clustering within the network.
Since R-MAT instances are apriori no flow networks, we choose the node with the largest outdegree as source and the one with the largest indegree as sink.
All arcs which are directly connected to the source and sink are safe from interdiction. The arc capacities are chosen independently at random from a uniform distribution on the interval $(0,1)$.

\sectionheadline{Results}

\begin{table}[bt]
	\centering
	\begin{widetable}{\textwidth}{lc|cccccccc}
		\toprule
		Network  & k & Iterations & \stackcell{Paths}{active/generated} & \stackcell{Pricing}{time} &  \stackcell{Scenarios}{active/generated} & \stackcell{Interdiction}{time} & \stackcell{Heuristic}{Gap} & \stackcell{Interdiction}{Gap} \\ \midrule
		P1 & \stackcell{5}{15}  & \stackcell{3}{5} & \stackcell{961/9636}{964/9597} & \stackcell{0.05s}{0.05s} & \stackcell{4/27}{20/42} & \stackcell{1.74s}{1.83s} & \stackcell{0}{0} & \stackcell{5.26\%}{5.26\%} \\ \midrule[0pt]
		P2 & \stackcell{5}{15} & \stackcell{3}{3} & \stackcell{97/459}{79/497} & \stackcell{0.04s}{0.24s} & \stackcell{1/97}{1/4509} & \stackcell{0.08s}{0.14s} & \stackcell{0}{0} & \stackcell{0}{0} \\ \midrule[0pt]
		P3 & \stackcell{5}{15} & \stackcell{7}{8} & \stackcell{73/1452}{86/924} & \stackcell{12.0s}{6.65s} & \stackcell{19/499}{20/4281} & \stackcell{0.30s}{3.32s} & \stackcell{1.32\%}{12.3\%} & \stackcell{25\%}{233\%} \\ \midrule[0pt]
		NETGEN-a & \stackcell{5}{15}  & \stackcell{4}{4} & \stackcell{82/181}{90/224} & \stackcell{0.50s}{3.17s} & \stackcell{1/708}{8/5638} & \stackcell{0.10s}{0.10s} & \stackcell{0}{0} & \stackcell{0}{3.75\%} \\ \midrule[0pt]
		NETGEN-b & \stackcell{5}{15} & \stackcell{3}{5} & \stackcell{97/181}{110/236} & \stackcell{0.09s}{2.52s} & \stackcell{1/180}{2/5413} & \stackcell{0.35s}{0.35s} & \stackcell{0}{0} & \stackcell{0}{0.92\%} \\ \midrule[0pt]
		R-MAT-a & \stackcell{5}{15} & \stackcell{4}{8} & \stackcell{88/140}{122/255} & \stackcell{0.08s}{1.57s} & \stackcell{4/180}{1/2049} & \stackcell{0.19s}{0.18s} & \stackcell{0}{0} & \stackcell{0.25\%}{0} \\ \midrule[0pt]
		R-MAT-b & \stackcell{5}{15} & \stackcell{4}{7} & \stackcell{148/231}{157/293} & \stackcell{0.14s}{8.02s} & \stackcell{1/85}{3/4798} & \stackcell{1.08s}{0.53s} & \stackcell{0}{0} & \stackcell{0}{0.12\%} \\ \midrule[0pt]
		R-MAT-c & \stackcell{5}{15} & \stackcell{4}{5} & \stackcell{67/168}{85/212} & \stackcell{0.27s}{3.64s} & \stackcell{1/120}{16/1224} & \stackcell{1.05s}{1.09s} & \stackcell{0}{3.24\%} & \stackcell{0}{0.53\%} \\ \bottomrule
	\end{widetable}
	
	\caption{Computational results for various network instances. Maximum of $3$ random samples. Times are $90\%$ quantiles of all pricing/interdiction calls.}
	\label{tab:results}
\end{table}

The separation procedure for the robust flow problem is implemented using the \cite{MATLAB:2015} interface of GUROBI (Version 6.5.0) to solve all occurring linear and mixed integer problems.
Computations are performed on a standard mobile workstation.
Hence, the reader is advised to focus on the number of iterations and variables and constraints generated instead of the actual computational time in seconds.
Table \ref{tab:results} summarizes the results for the procedure as described in Section~\ref{sec:solution-techniques}.
The randomized instances are investigated for $3$ different seeds, and the maximum values are reported.

The listed times for the pricing and interdiction problems refer to the $90\%$ quantile of all MIP solutions which are computed until optimality is achieved.
As the bounded integrality gap of model (\ref{lpseparation}) suggests, the interdiction problem can be solved fast.
The pricing problem turns out to be harder, yet uncritical, since the number of generated paths is significantly lower than the set of scenarios. The higher computation time for the pricing problem may also be attributed to the larger number of scenarios which increase the size of the pricing problem.
Thus, the primary source of complexity is the number of scenarios to be generated, rather than the generation itself.
This reflects the original issue which motivated our computational technique, namely the huge size of the linear program (\ref{lp:robustfull}).

Since the separation procedure allows to solve the robust flow problem optimally, we also evaluate the performance of the heuristic suggested by \cite{bertsimas2013robust}.
For most of the considered instances, the heuristic solution is optimal. It is an open question to characterize the networks for which this optimality holds.
\cite[Corollary 4.1]{baffier2014parametric} present a criterion which implies that the heuristic value equals the value of the network interdiction problem for the full graph as studied by \cite{wood1993deterministic}.
In general, the value of the network interdiction problem is an upper bound on the robust flow value. Table \ref{tab:results} reports the gap between the optimal solution and this upper bound.
Since the gap is positive for most graphs, we conclude that the criterion of Baffier et al.\ does not apply here.

\begin{figure}[bt]
	\centering
	\includegraphics[page=1]{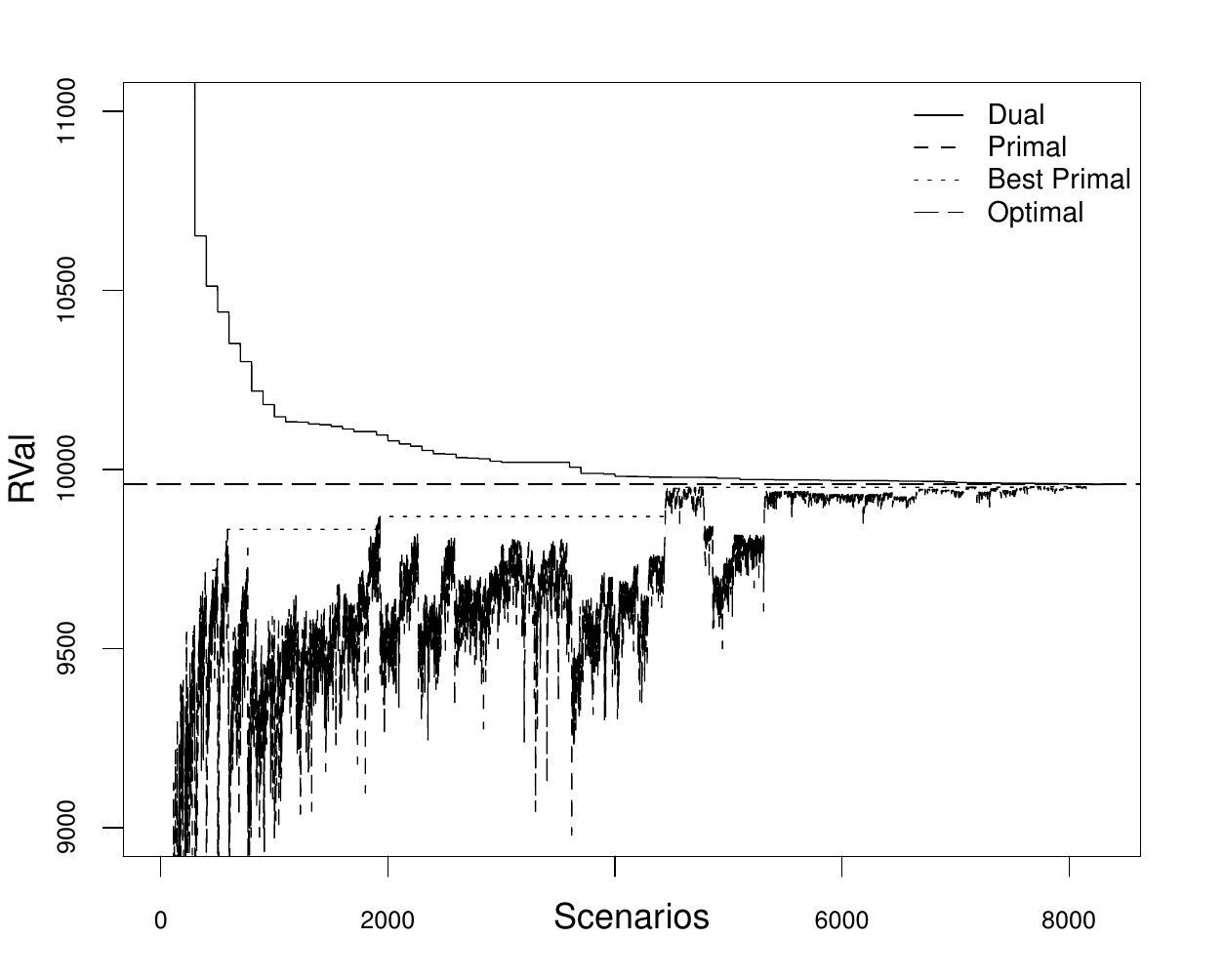}
	\caption{Trend of the primal and dual bound with a restricted count of interdictions per iteration for a NETGEN-b instance. Top: 10 interdictions; Bottom: 100 interdictions.}
	\label{fig:iterations}
\end{figure}

\begin{figure}[bt]
	\centering
	\includegraphics[page=1]{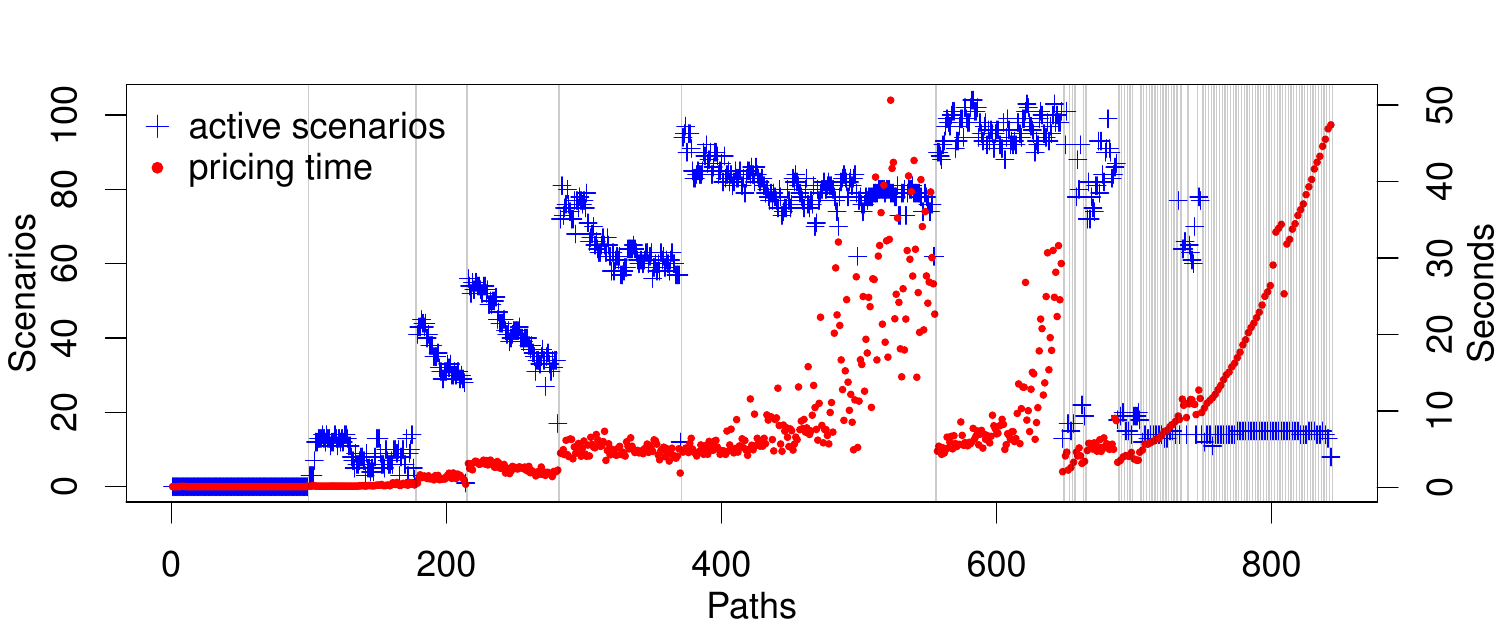}
	\includegraphics[page=1]{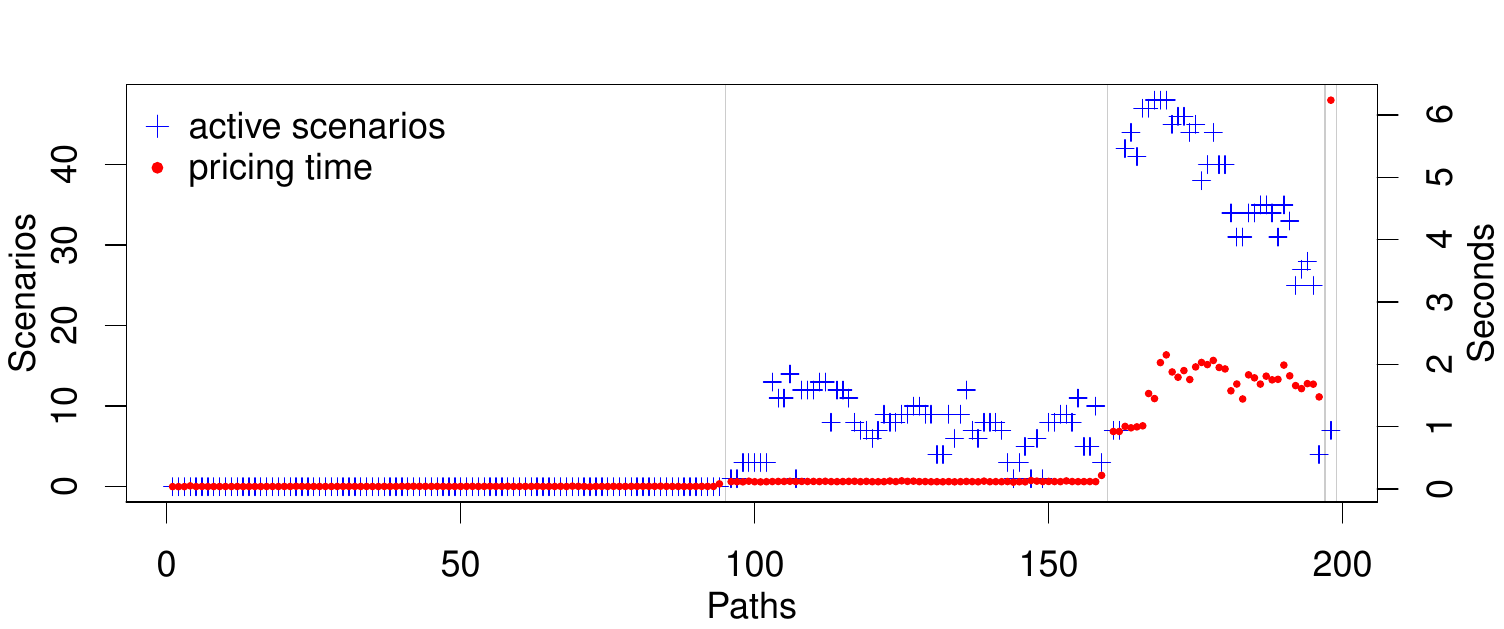}
	\caption{Positive scenario cost coefficients and MIP solution time for each call of the pricing problem, performed for a NETGEN-a instance. Vertical lines indicate the end of a primal-dual iteration cycle. Top: 100 interdictions per iteration; Bottom: complete interdiction.}
	\label{fig:pricing-times}
\end{figure}

The highest count of scenarios has been generated for instances of type NETGEN-a.
For this specific network, we run the alternative procedure outlined in Section \ref{sec:solution-techniques} by restricting the number of scenarios generated per iteration to 100.
That is, each iteration of the procedure consists of pricing all required paths into the problem to achieve optimality for the current set of scenarios, and then generating 100 new interdiction scenarios trying to improve the primal bound.
After generating approximately 8200 scenarios and about 800 paths in this manner, the procedure achieves an optimality gap of $10^{-4}$.
For comparison, leaving the number of generated interdiction scenarios per iteration unrestricted, the algorithm terminated with 5600 scenarios and about 200 paths (Table \ref{tab:results}).
For the restricted procedure, Figure \ref{fig:iterations} depicts the trend of the primal and dual bounds as the set of scenarios grows.
As can be seen, most calls of the interdiction problem (\ref{lpseparation}) lead to values which do not improve the primal bound.
In contrast, the standard approach of adding all relevant interdiction scenarios to the linear program is guaranteed to terminate with a solution which is at least as good as the primal bound from the previous iteration.

Figure \ref{fig:pricing-times} depicts the (empirical) complexity of the encountered pricing problems.
We find that the solution time of the pricing problem is significantly higher if the number of interdictions is restricted.
A possible source of complexity might be the structure of the cost vector for the pricing problem.
Yet, comparing the two approaches, the sparsity resp.\ denseness of the price vector is roughly the same for the restricted and unrestricted case.
Hence the increased difficulty of the pricing problem for this case can only be attributed to more complex properties of the polyhedron.

\section{A stochastic model for maximum flows}
\label{sec:stochastic-model}

One major motivation for the study of robust flows in particular, and robust optimization in general, is the requirement to deal with uncertain data.
Since uncertainty is typically described by methods of probability theory, a stochastic formulation seems to be more natural for this task.
As shown in this section, both approaches turn out to be closely related for a wide range of problems including robust flows.

In a general setting, let $X\subset \R^n$ denote the solution space of a given problem, and $\Omega$ be an arbitrary set of possible scenarios.
If a scenario $\omega\in\Omega$ occurs, the value of a given solution $x\in X$ is described by the payoff function $\val_\omega(x)$.
In a stochastic context, we are given a probability measure $Q$ on the scenario set $\Omega$. Stochastic optimization aims to determine a feasible solution $x$ which maximizes the expected payoff, that is,
\begin{align}
\label{stochobj}\max_{x\in X} E (\val_\omega(x))=\int_\Omega \val_\omega(x) \, dQ(\omega).
\end{align}
In the context of robust maximum flows, $X$ describes the set of feasible path-flows, and each $\omega$ is a set of edges that may be destroyed simultaneously, $\Omega=2^E$.
The value function is the remaining flow after removal of $\omega \subseteq E$, that is, $\val_\omega(x) = \sum_{P\in\mathcal{P}: P \cap \omega = \emptyset} x_P$.
Thus, solving (\ref{stochobj}) is typically intractable since it requires a representation of the value function $\val_\omega$ for an exponential number of scenarios $\omega \in \Omega$.
Robust models can be interpreted as a computational remedy: instead of considering all scenarios $\omega$, only the worst case is of interest.
As shown in the previous sections, this is often possible by considering only a fraction of the full scenario set.

The increase in computational tractability comes at the price of modeling flexibility.
To regain it in part, we suggest the following hybrid approach.
For $K\in\mathbb{Z}_+$, let $\Omega_1,\ldots,\Omega_K$ be a partition of $\Omega$, and specify probabilities $q_k=Q(\Omega_k)$, that is, $q_k\geq 0$ such that $\sum_{k=1}^K q_k=1$.
Instead of (\ref{stochobj}), we may consider the relaxed model \begin{align}
\label{stochrobobj}\max_{x\in X} \sum_{k=1}^K q_k \min_{\omega\in\Omega_k} \val_\omega(x).
\end{align}
Instead of a probability for each event individually, we aggregate sets of events and assume that, in each partition, the worst-case occurs.
Hence, the optimal value of (\ref{stochrobobj}) always yields a lower bound on the optimal value of (\ref{stochobj}).

In case of the maximum flow problem, the partition $\Omega_k = \binom{E}{k}$ is a natural candidate.
That is, the events of cardinality $k$ are aggregated into a single event which is counted by its worst-case impact.
Setting $q_k = 1$ for some $k$ corresponds to the robust flow problem from the previous sections, whereas the full stochastic model (\ref{stochobj}) may be recovered by $K = |\Omega|$, which implies that each $\Omega_k$ contains a single scenario.
Interestingly, the decomposition techniques discussed in Section \ref{sec:solution-techniques} extend to model (\ref{stochrobobj}). For $\Omega_k=\binom{E}{k}$, the primal separation problem is solved for each $k$ individually, and the pricing problem can be adjusted accordingly by adding additional variables $z_\eta$.

One might ask if the aggregation of scenarios introduces a large gap in optimal objective function values between models (\ref{stochrobobj}) and (\ref{stochobj}).
We can show that the lower bound of the hybrid approach is tight in the following sense.
For each specification of $(\Omega_k, q_k)_{k=1}^K$, there is a probability distribution $Q$ on $\Omega$ satisfying $Q(\Omega_k) = q_k$, denoted $Q\equiv q$, such that the solution of (\ref{stochrobobj}) is optimal for the stochastic model (\ref{stochobj}).

\begin{theorem}\label{thm:stochrobust}
	Let $X$ be a compact, convex subset of $\R^p$, $\Omega$ be an arbitrary set and $\Omega_k\subset \Omega, k=1,\ldots, K$ be a partition of $\Omega$.
	If $x\mapsto \val_\omega(x)$ is continuous and concave for each $\omega\in\Omega$, then
	\begin{align*}
	\max_{x\in X} \sum_{k=1}^K q_k \inf_{\omega\in\Omega_k} \val_{\omega}(x) = \inf_{Q\equiv q} \max_x \int \val_\omega(x)\, dQ(\omega).
	\end{align*}
\end{theorem}
\proof{Proof.}
The set $\{ Q: Q\equiv q \}$ is a linear subspace of probability measures on $\Omega$.
By linearity of the integral, $E_Q(\val_\omega(x))$ is linear in $Q$, concave in $x\in X$ and, by Fatou's Lemma, upper semi-continuous in $x$ (see e.g.\ \cite{billingsley2013convergence}).
Thus, an application of a classical min-max theorem (\cite{fan1953minimax}) yields
\begin{align}
\label{eqnminmax}\inf_{Q\equiv q} \max_{x\in X} E_Q(\val_\omega(x)) = \max_{x\in X} \inf_{Q\equiv q} E_Q(\val_\omega(x)) \geq \max_{x\in X} \sum_{k=1}^K q_k \inf_{\omega\in\Omega_k} \val_{\omega}(x).
\end{align}
In order to show equality in (\ref{eqnminmax}), let $\omega_k^n\in\Omega_k$ be a minimizing sequence of $\val_\omega(x)$, not necessarily convergent.
Let us consider the following probability measure $Q^n=\sum_{k=1}^K q_k \delta_{\omega_k^n}$, where $\delta_\omega$ is a point mass at $\omega\in\Omega$.
This measure satisfies $Q^n(\Omega_j)=\sum q_k \delta_{\omega_k^n}(\Omega_j)=q_j$ for each set $\Omega_j$.
Hence, $Q^n\equiv q$.

By choice of the sequence $\omega_k^n$, integration with respect to $Q^n$ yields
$$E_{Q^n}(\val_\omega(x))= \sum_{k=1}^K q_k \val_{\omega_k^n}(x)\to \sum_{k=1}^K q_k \inf_{\omega\in\Omega_k} \val_\omega(x) \text{, as } n\to\infty.$$
Thus, $Q^n$ is a minimizing sequence for $\inf_{Q\equiv q} \max_{x\in X} E_Q(\val_\omega(x))$ which attains the right hand side of (\ref{eqnminmax}). \Halmos
\endproof

The robust maximum flow model fits into Theorem \ref{thm:stochrobust} since the set of feasible flows forms a convex polyhedron.
The scenario set $\Omega=\{\eta \subseteq E : |\eta| \leq k\}$ is a discrete topological space, and the function $x\mapsto \val_\eta(x)$ is linear for each scenario $\eta\in\Omega$.

While (\ref{stochrobobj}) may be motivated as a computational heuristic for the full stochastic model (\ref{stochobj}), it is also of interest if the full model is tractable.
Since an instance for (\ref{stochrobobj}) requires the values of a probability distribution $Q$ only for each set $\Omega_k$, it may be less prone to misspecifications of the uncertainty set.
In this sense, it can be seen as a trade-off between the flexibility of a stochastic model and the conservativeness of a robust solution.

The stability with respect to the probability measure $Q$ is of particular interest since, in practice, $Q$ is not known but has to be estimated from historical data and forecasts.
For example, applying the robust flow model on a daily basis, the probability $Q(\eta)$ for $\eta \subseteq E$ describes the probability of arcs in $\eta$ failing on any given day.
Given past observations $\eta_1,\ldots, \eta_N$, it may be estimated by the empirical distribution
\begin{align*}
\hat{Q}(\eta)=\frac{1}{N} \sum_{i=1}^N \textbf{1}_{\eta=\eta_i} =\; \textit{"fraction of days on which exactly arcs in $\eta$ failed"}.
\end{align*}
Since the number of scenarios $\eta \subseteq E$ is of exponential size, the historical record has usually not witnessed every scenario.
Hence, $\hat{Q}(\eta) = 0$ for most scenarios $\eta$ although the true probability $Q(\eta)$ is likely to be positive.
Thus, an optimal flow would avoid scenarios $\eta$ which have previously occurred and underestimate the true risk.
This general phenomenon is well known in the literature on statistical learning theory and commonly referred to as generalization error or overfitting (e.g.\ \cite{bousquet2002stability}).

Considering only the aggregated probabilities $\hat{Q}(\Omega_k) = \sum_{\eta\in \Omega_k}\hat{Q}(\eta)$ generalizes the historical observations towards classes of events which are assumed to occur with equal probability.
The larger the sets $\Omega_k$, the higher the degree of generalization. This reasoning is made precise by the following Rademacher complexity bound.
For the maximum flow problem, Theorem \ref{thm:rademacher} estimates the performance of the hybrid flow in terms of its empirical performance. The bound gets stronger as the number $N$ of historically observed events grows.

\begin{theorem}\label{thm:rademacher}
	Let $G=(E,V)$ be a directed graph with arc capacities $(u_e)_{e\in E}$, $\lambda_k^*=\max_{\eta\in \Omega_k}\sum_{e\in\eta} u_e$ and $C=\max_k \lambda_k^*$.
	Assume that the scenarios $\eta_1,\ldots, \eta_N \in \Omega$ are drawn independently from $Q$ (The joint probability will be denoted by $Q$ as well).
	Then, for any $\delta\in(0,1)$ and for any feasible s-t-flow $(x_P)_{P\in\mathcal{P}} \in \mathcal{F}$, the following holds with probability at least $1-\delta$:
	\begin{align*}
	\sum_{k=1}^K q_k \min_{\eta\in\Omega_k} \val_\eta(x) &\geq  \sum_{k=1}^K \hat{q}_k \min_{\eta\in\Omega_k} \val_\eta(x) \\
	&\qquad - \frac{2}{\sqrt{N}}\sum \lambda_k^* \sqrt{\hat{q}_k} - 2C\sqrt{ \frac{8\ln(4/\delta)}{N} }.
	\end{align*}
\end{theorem}

\proof{Proof.}
For each $k$, let $\lambda_k(x)$ describe the amount of flow destroyed by the worst-case scenario in $\Omega_k$.
That is, $\lambda_k(x)=\max_{\eta\in\Omega_k}\sum_{P\cap\eta\neq \emptyset} x_P$ such that $\min_{\eta\in\Omega_k} \val_\eta(x)=\sum_P x_P - \lambda_k(x)$.
The vector $\lambda(x)$ satisfies the bound $\lambda_k(x)\leq \max_{\eta\in\Omega_k} \sum_{e\in\eta} u_e=\lambda_k^*$.
Define $X_i\in\{0,1\}^K$ such that $X_i^k=1$ if and only if $\eta_i\in\Omega_k$.
Then $\hat{q}=\frac{1}{N}\sum_{i=1}^N X_i$ componentwise, and $E_Q(\hat{q})=q$. Hence, the discrepancy between the true and the estimated performance may be written as
\begin{align*}
\sum_{k=1}^K {q}_k \min_{\eta\in\Omega_k} \val_\eta(x) - \sum_{k=1}^K \hat{q}_k \min_{\eta\in\Omega_k} \val_\eta(x) &= \sum_{k=1}^K (\hat{q}_k - q_k)\lambda_k(x)\\
&= \sum_{k=1}^K \left(\frac{1}{N} \sum_{i=1}^N X_i^k -E(X_1^k)\right) \lambda_k(x)\\
&= \frac{1}{N}\sum_{i=1}^N \lambda(x)^T X_i -E_Q(\lambda(x)^T X_1).
\end{align*}
The quantity $\lambda(x)^T X_i$ is upper bounded by $C$.
It may be related to its expectation in a uniform fashion by means of a Rademacher complexity bound.
\cite{bartlett2002rademacher} show that \begin{align*}
\frac{1}{N} \sum_{i=1}^N \lambda(x)^T X_i \geq E_Q(\lambda(x)^T X_1) - E_Q(\hat{R}_N) - C\sqrt{\frac{8\ln(4/\delta)}{N}},
\end{align*}
holds with probability at least $1-\delta/2$,
where \begin{align}
\label{defrademacher}\hat{R}_N = E_\sigma \sup_{\lambda: 0\leq \lambda_k \leq \lambda_k^*} \frac{2}{N} \left| \sum \sigma_i (\lambda^T X_i) \right|
\end{align}
is the empirical Rademacher complexity (see Theorem 8). The expectation in (\ref{defrademacher}) is taken only with respect to $\sigma_i$, which is a sequence of Rademacher random variables independent of $X_i$. That is, $Q(\sigma_i=1)=Q(\sigma_i=-1)=\frac{1}{2}$. An application of Theorem 11 of \cite{bartlett2002rademacher} with $\epsilon=C\sqrt{{8\ln(4/\delta)}/{N}}$ yields \begin{align*}
\frac{1}{N} \sum_{i=1}^N \lambda(x)^T X_i \geq E_Q(\lambda(x)^T X_1) - \hat{R}_N - 2C\sqrt{\frac{8\ln(4/\delta)}{N}},
\end{align*}
with probability at least $1-\delta$.

The Rademacher complexity $\hat{R}_N$ measures the ability of the vectors $\lambda$ to fit random noise.
Due to the binary nature of the variables $X_i$, we may compute the bound explicitly:
\begin{align*}
\hat{R}_N &= E_\sigma\sup_{\lambda: 0\leq \lambda_k \leq \lambda_k^*} \frac{2}{N}\left|  \sum_{k=1}^K \lambda_k \sum_{i: \eta_i\in \Omega_k} \sigma_i   \right|\\
&\leq E_\sigma \frac{2}{N}  \sum_{k=1}^K \lambda_k^* \left|\sum_{i: \eta_i\in \Omega_k} \sigma_i   \right|\\
&\leq \frac{2}{N}  \sum_{k=1}^K \lambda_k^* \sqrt{E_\sigma\left(\sum_{i: \eta_i\in \Omega_k} \sigma_i   \right)^2}.
\end{align*}
The sum $\sum_{i: \eta_i\in\Omega_k} \sigma_i$ consists of $N \hat{q}_k$ independent summands of unit variance, hence \begin{align*}
\hat{R}_N &\leq \frac{2}{N} \sum_{k=1}^K \lambda_k^* \sqrt{ N \hat{q}_k } = \frac{2}{\sqrt{N}} \sum_{k=1}^K \lambda_k^* \sqrt{\hat{q}_k },
\end{align*}
which completes the proof. \Halmos
\endproof

As Theorem \ref{thm:rademacher} demonstrates, the capability of the hybrid model to generalize from past observations is approximately determined by the term $\sum_{k=1}^K \sqrt{\hat{q}_k}$.
Since the square-root is sub-additive, this bound is smaller the larger each individual value $\hat{q}_k$ is.
Summarizing scenarios into larger sets $\Omega_k$ thus prevents the optimization model from overfitting to historical data.

By choosing the granularity of the partition $\{\Omega_k\}$, a compromise needs to be found between a detailed model which might underestimate the true risk and a more pessimistic model of increased reliability.
The lower bound of Theorem \ref{thm:rademacher} can be used to compare different models based on their empirical performance, as the bound explicitly corrects for the complexity of the model.

In the setting of robust flows, the sets $\Omega_k$ may naturally be chosen to contain all attacks of $k$ arcs, that is, $\Omega_k=\binom{E}{k}$.
To obtain a more detailed model, the set of arcs may be partitioned into regular and exposed arcs, $E = E_1 \cup E_2$.
The scenarios could be partitioned as $\Omega_{j,k}=\{ \eta\subset E : |\eta\cap E_1|=j, |\eta\cap E_2|=k \}$ of $j$ attacks on regular and $k$ attacks on exposed arcs.

\section{Conclusion}
\label{sec:conclusion}

We presented an approach to handle path-based robust flows computationally.
By alternating primal and dual separations, the size of the corresponding linear program remains tractable while yielding upper and lower bounds on the optimal solution value.
A simulation study for different types of networks revealed that the NP-hard separation problem can be solved by a standard branch-and-bound procedure in practice.
Furthermore, the results verify optimality of an heuristic solutions for most of the considered instances.

Furthermore, we introduced a new model for path-based robust flows which combines robust and stochastic aspects.
The model can be treated computationally by the approach presented in this paper.
We were able to describe the balance between modeling flexibility and robustness by means of a bound on the generalization error when the model parameters are based on historical data.

The new model as well as the computational approach can in principle be extended to other planning problems under uncertainty.
However, both results rely on convexity of the solution space.
Thus, a straight-forward extension to integral problems is not conceivable.

\addcontentsline{toc}{chapter}{Bibliography}
\Urlmuskip=0mu plus 1mu
\bibliographystyle{apalike}
\bibliography{robust-flow-comp}

\end{document}